\documentclass[a4paper]{article}
\usepackage{icrc2013}
\usepackage[english]{babel}
\usepackage{amssymb,amsmath}

\usepackage{auto-pst-pdf}
\usepackage{epstopdf}
\usepackage{epsfig}

\def\Journal#1#2#3#4{{#4}, {#1}, {#2}, #3} 

\title{Identification of Light Cosmic-Ray Nuclei with AMS-02} 
\shorttitle{Light-Nuclei ID with AMS-02}                      

\authors{
N. Tomassetti$^{1,2}$ \& A. Oliva$^{3}$, for the AMS Collaboration.
}

\afiliations{
$^{1}$ LPSC, Universit\'{e} Joseph Fourier Grenoble 1, CNRS/IN2P3, Institut Polytechnique de Grenoble, F-38026 Grenoble, France. \\
$^{2}$ INFN-Sezione di Perugia, I-06100 Perugia, Italy. \\
$^{3}$ Centro de Investigaciones Energeticas, Medioambientales y Tecnologicas, CIEMAT, E-28040 Madrid, Spain. \\
}

\email{nicola.tomassetti@lpsc.in2p3.fr}
\abstract{
AMS-02 is a wide acceptance (0.5\,m$^2$ sr) and long duration (up to 20 years) magnetic spectrometer 
operating onboard the International Space Station since May 2011. Its main scientific objectives are the indirect research 
of Dark Matter, searches of primitive Anti-Matter and the precise measurement of the Cosmic-Ray (CR) spectra. 
Among charged CR species, AMS-02 will be able to measure relative abundances and absolute fluxes of CRs nuclei 
from Hydrogen up to at least Iron ($Z=26$) in a kinetic energy range from hundreds MeV to TeV per nucleon. The high 
statistics measurement of the chemical composition of CRs in this extended energy range will reveal new insights 
about the CRs life in the Galaxy, from their origin to the propagation in the interstellar 
medium, giving new constraints to astrophysical models of Galactic CRs.
The nucleus absolute charge, $Z$, is measured several times along the trajectory of the particle inside AMS-02 using 
different detection techniques: in the 9 planes of the Silicon Tracker, in the 4 layers of scintillator counters of the 
Time-of-Flight system (TOF), in the Ring Imaging Cherenkov Counter (RICH) as well as in the 20 layers of Transition 
Radiation Detector (TRD) and in the upper layers of the Electromagnetic Calorimeter (ECAL).
The combination of the redundant measurements delivered by the tracking system and by the TOF allows an accurate 
discrimination between chemical elements. The charge measurements in the detectors on top of AMS, as the Upper plane 
of the Tracker and in TRD, is used for the identification of the incoming nuclear specie and allows the charge-changing 
events background estimation.
The AMS-02 different charge measurement principles are here briefly explained, and performance of each sub-detector presented. 
Then the AMS-02 combined charge separation capability as well as the interaction events identification principles are presented.
}


\begin{document}
\maketitle

\section{Introduction} 

Hadrons are the main component of the CR flux. In the  $\sim$\,GeV--TeV range of kinetic energy, CRs are 
composed mainly of protons (about $\sim$\,90\%) and He ($\sim$\,9\%), while heavier nuclei constitute $\sim$\,1\% of the flux.
Part of them such as p, $^{4}$He, C-N-O, or Fe, are believed to be of primary origin, i.e., accelerated by supernova remnant (SNR) explosions,
although the exact mechanisms are not yet well known. 
Rarer CR elements such as $^{2}$H, $^{3}$He and Li-Be-B are believed to be of secondary origin, i.e. produced by collisions primary 
CRs with the gas nuclei of the interstellar medium (ISM). The secondary CR flux depend on the abundance of their progenitors nuclei, 
their production rate and their diffusive transport in the ISM \cite{Bib::Strong2007,Bib::Maurin2001}.
Thus, secondary to primary ratios such as Li/C, B/C, or F/Ne are used to discriminate among astrophysical models of CR propagation in the Galaxy.
Furterly, it was recently pointed out that the B/C ratio at $\gtrsim$\,100\,GeV/n is crucial to understand 
open problems in CR physics such as the observed structures in primary CR spectra \cite{Bib::Vladimirov2012,Bib::Tomassetti2012}, 
or secondary production processes inside SNR shock waves \cite{Bib::Mertsch2009,Bib::Tomassetti2012b}.

The Alpha Magnetic Spectrometer (AMS) is an international project
devoted to study Galactic CRs by direct detection of sub-GeV -- TeV particles in space.
The main goals of the experiment are the direct search of anti-nuclei and
indirect search of dark matter particles through their annihilation
into light particles such as $\bar{p}$, $e^{\pm}$, or $\gamma$\--rays \cite{Bib::Aguilar2013}.
The final version of the detector, AMS-02, was successfully installed on the International Space Station
on May 19$^{\rm th}$ 2011. After its installation and activation, AMS-02 has recorded several billions  
CR particles, from $Z=1$ to $Z=26$, at energies from a few $\sim$\,100\,MeV to $\sim$1\,TeV per nucleon, 
with unprecedent precision and sensitivity.
The AMS-02 data are expected to significantly improve our
understanding of the CR acceleration and propagation processes in the Galaxy \cite{Bib::Oliva2008,Bib::Pato2010}.
The status of the ongoing analyses on light-nuclei is reviewed in these proceedings \cite{Bib::ICRC-AMSBC}.

\section{CR Nuclei Measurement with AMS-02} 
\label{Sec::AMS}                           

\begin{figure*}[!t]
  \centering
  \includegraphics[width=0.97\textwidth]{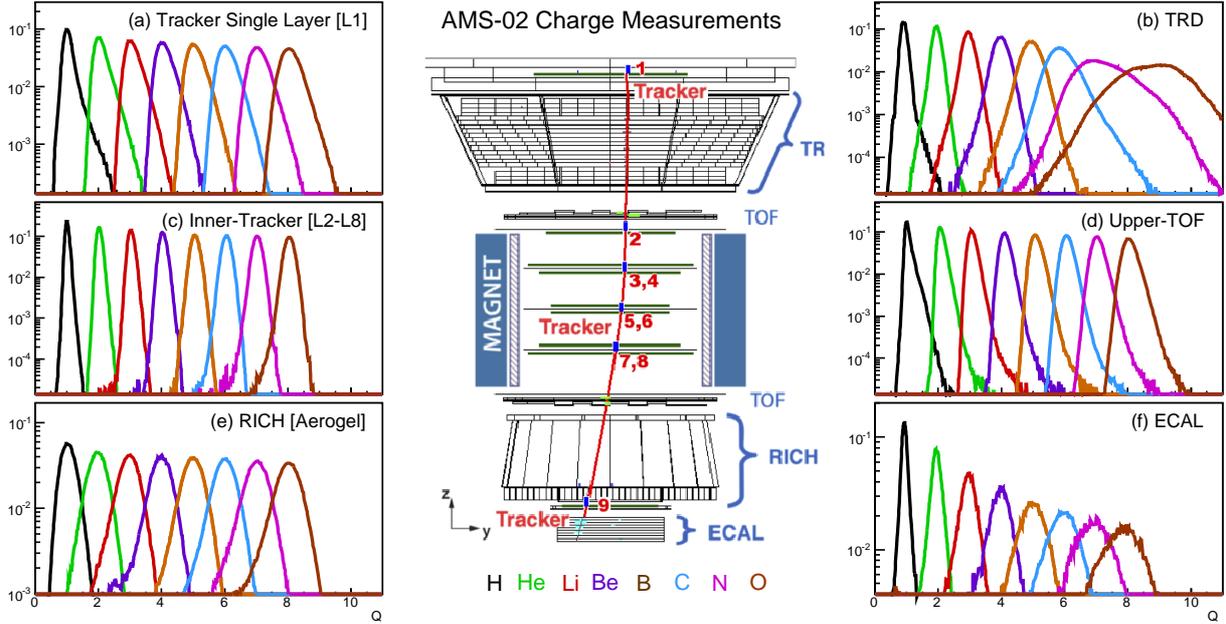}
  \caption{
    Schematic Y-Z view of the AMS-02 detector, illustrating the path of a typical CR event. 
    The panels show the charge response of single sub-detector units to light CR elements ($Z=1$ to $Z=8$): 
    (a-c) Tracker \cite{Bib::ICRC-Tracker}, (b) TRD \cite{Bib::ICRC-TRD}, (d) TOF \cite{Bib::ICRC-TOF}, 
    (e) RICH \cite{Bib::ICRC-RICH}, and (f) ECAL \cite{Bib::ICRC-ECAL}.}
  \label{Fig::ccAMSDetector}
 \end{figure*}

The AMS-02 instrument is sketched in Fig.\,\ref{Fig::ccAMSDetector}. The figure illustrates the 
basic principle of CR measurement for a typical event.
The main characteristics of a CR particle traversing AMS-02 are the arrival direction, 
the particle identity and its kinetic energy or rigidity (i.e., momentum to charge ratio $R=p/Z$).
These quantities are reconstructed in AMS-02 by combining independent measurements provided by the various sub-detectors.
The particle direction and rigidity are obtained by 
the reconstruction of its trajectory along up to nine Silicon Tracker layers
with $\sim$\,10\,$\mu$m ($\sim$\,30\,$\mu$m) of spatial resolution on the Y (X) side.
\cite{Bib::Tracker}.
The velocity $\beta = v/c$ can be determined from the transit time between the upper-TOF and lower TOF 
scintillator planes along the track (for $Z=1$, $\Delta\beta/\beta\,\sim\,3\%$ \cite{Bib::TOF}), or more precisely using the 
RICH system (for $Z=1$, $\Delta\beta/\beta\,\sim\,10^{-3}$ \cite{Bib::RICH}).
The CR elemental species is specified by the reconstruction of the CR nuclear charge $Z$, 
which is crucial task in the data analysis of CR spectra. This will be mostly MC driven, although
some templates (at least for non-interacting events) can be also be built with the help of data.

\section{Charge Measurements} 
\label{Sec::ChargeMeasurements} 

Information on the CR nuclear charge $Z$ are obtained by the multiple measurements of energy loss
in the several detectors of the spectrometer and by the amount of Cherenkov light detected by the RICH. 
Details on the measurement principle and on the calibration procedures adopted for the charge identification 
can be found in these proceedings \cite{Bib::ICRC-TOF,Bib::ICRC-TRD,Bib::ICRC-Tracker,Bib::ICRC-RICH,Bib::ICRC-ECAL}.
Following the trajectory within the instrument (see Fig.\,\ref{Fig::ccAMSDetector}) charged CR particles
traverse up to 9 double-sided Tracker layers of micro-strip silicon sensors, 20 TRD layers \cite{Bib::TRD},
and 2+2 TOF planes before emitting Cherenkov radiation in the RICH counter and eventually 
get absorbed into the 18 ECAL layers \cite{Bib::ECAL}.

The adopted analysis strategy is to identify the CR nuclear charge using the 7 Inner-Tracker (IT) layers (from L2 to L8)
and the 4 TOF planes, while complementary information from the other sub-detectors are used in a second stage.
The IT+TOF combination represents the basic core of the AMS-02 spectrometer and ensures 
charge separation capabilities over a wide dynamical range ($Z=1 - 26$).
Moreover, the particles paths within the IT tracking volume traverse a relatively small amount of material ($\sim$\,1.5\,g/cm$^{2}$).
Interactions within the material occurs mostly in the upper or lower part of the spectrometer
and are studied by means of redundant charge measurements.
In fact, using the outer Tracker layers (L1 and L9), TRD, RICH or ECAL one can detect
the appearance of fragmentation processes at different levels in the spectrometer (\S\ref{Sec::NuclearInteractions}). 

In the IT layers, the ionization energy generated by a charged particle in 
the silicon sensors is recorded as two paired clusters of readout micro-strips, 
on both sides, giving 2-dimensional hits. Tracker clusters are recognized online and then reprocessed 
by the reconstruction software. The ADC values of the readout cluster amplitudes
$A$ is related to the energy deposition $\frac{dE}{dx}$ which, in turn, depends on the
CR nuclear charge and velocity. The most probable energy deposition is roughly
$\langle \frac{dE}{dx} \rangle \propto \frac{Z^{2}}{\beta^{2}}\log{\gamma}$,
while each energy deposition is Landau-distributed around this value.
The relation between the cluster amplitudes $A$ and energy loss $\frac{dE}{dx}$ is influenced by 
charge collection inefficiencies, signal attenuation and deviations from linearity. Thus, a multi-step procedure 
of signal linearization and normalization has been performed as a function of impact position $\eta$ and inclination 
$\theta$ of the particle on the sensors, as well as chip-based corrections for electronic response \cite{Bib::ICRC-Tracker}.
Finally, a maximum-likelihood $Z$\--estimator has been developed to combine the corrected cluster 
signals $q_{j}=q_{j}(A,\beta, \theta,\eta)$, here expressed in charge units,
to determine the ``best'' nuclear charge $Z$ associated with observations. 
The IT likelihood function is defined as:
\begin{equation}
\mathcal{L}_{Z} = \prod_{j=1}^{j=N_{2D}}P_{Z}(q^{X}_{j},q^{Y}_{j}) \times \prod_{j=1}^{j=N_{X}}P_{Z}(q^{X}_{j}) \times \prod_{j=1}^{j=N_{Y}}P_{Z}(q^{Y}_{j}) \,,
\end{equation}\label{Eq::Likelihood}
where $q^{X}_{j}$ and $q^{Y}_{j}$ are the corrected cluster amplitudes for the $j$\--th hit on the $X$ and $Y$ sides, 
and $P_{Z}$ are the corresponding probability density functions (PDF) for $Z$\--charged particles. 
The likelihood function accounts for the joint probabilities of $N_{2D}$ 2-dimensional hits ($P_{z}(q^{X},q^{Y})$) 
as well as $N_{X}$ and $N_{Y}$ one-dimensional clusters (when they are unpaired) on both sides.
The PDFs are evaluated layer by layer using kernel density estimators of clean data samples with known $Z$ particles. 
PDFs of the first tracker layer are shown in Fig.\,\ref{Fig::ccAMSDetector}a for $Z=1$ to $Z=8$. Other Tracker layers have similar charge responses. 
The best IT charge is defined as the $Z$ value for which the geometric mean likelihood $\mathcal{H}_{Z}=\sqrt[N]{\mathcal{L}_{Z}}$
takes the maximum value, where $N=2\cdot N_{2D}+N_{X}+N_{Z}$. 
The above procedure allows to assign integer estimators $Z$ to all CR events traversing the AMS-02 acceptance. 
It is useful to define a global charge estimator $Q_{\rm IT}$ (floating value), which we define as
combined $X-Y$ mean of all corrected cluster amplitudes $q_{j}$. 
 \begin{figure}[!t]
  \centering
  \includegraphics[width=0.525\textwidth]{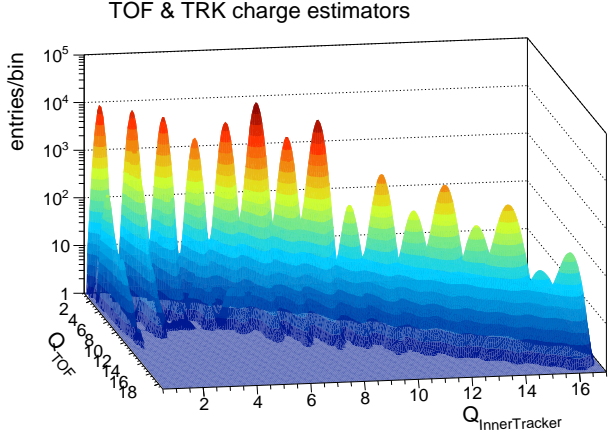}
  \caption{$Q_{\rm IT}$ VS $Q_{TOF}$ floating charge estimators for $Z=1$ to $Z=16$.
    For graphical clarity, p and He nuclei have been rescaled factors 1/1000 and 1/200, respectively.}
  \label{Fig::ccChargeTrkVSTof}
 \end{figure}
Similarly, a TOF charge estimator $Q_{\rm TOF}$ has been defined as the truncated mean of the CR energy depositions 
in the four TOF planes (following the TOF reconstructed track), after corrections for velocity, energy loss, 
attenuation, impact point, impact angle, and readout effects \cite{Bib::ICRC-TOF}. 
For each $Z$, the quantities $Q_{\rm IT}$ and $Q_{\rm TOF}$ are gaussian-like distributed and their widths
give the IT charge separation power, namely the charge resolution $\delta Z$ \cite{Bib::ICRC-Tracker}.
With the IT one obtains $\delta Z\lesssim$\,0.1 up to Carbon and $\lesssim\,0.3$ up to Silicon,
following an approximate behavior $\delta Z\,\sim\,0.04\times Z + 0.02$.
Consequence of the finite charge resolution, arising from fluctuations in the energy depositions,
is the charge mis-identification which represents one source of background for CR nuclei measurement.
The IT alone has excellent identification capabilities, giving 
$P(Z_{\rm IT}\neq Z|Z) \lesssim$\,10$^{-4}$ for light nuclei up to Oxygen \cite{Bib::ICRC-Tracker}.
Figure\,\ref{Fig::ccChargeTrkVSTof} illustrates the distribution of $Q_{\rm TOF}$ VS $Q_{\rm IT}$ for relativistic 
($\beta_{TOF}>0.95$) CRs nuclei with from $Z=1$ (protons) up to $Z=16$ (Sulfur). 
The correlation of the two independent charge measurement is apparent.
The single charge estimators are shown in Fig.\,\ref{Fig::ccAMSDetector}c for the IT and 
Fig.\,\ref{Fig::ccAMSDetector}d for the Upper-TOF planes \cite{Bib::ICRC-Tracker,Bib::ICRC-TOF}. 
The Lower-TOF counters give a similar charge response. 
The other panels show analogue charge estimators obtained with other detector units.
Note that each plot of Fig.\,\ref{Fig::ccAMSDetector} is referred to a specific sub-detector where 
the various CR elements ($Z=1\--8$) have been identified using information from other sub-detectors. 
Each colored curve of the figure represents the a charge PDF, i.e., the charge response of each 
sub-detector units to $Z$\--charged CR nuclei, expressed in charge units and normalized to unit area.

\section{Nuclear Interactions}   
\label{Sec::NuclearInteractions} 

The presence of several charge detectors above and below the IT-TOF core of the spectrometer
are be used to improve the overall charge separation capability as well as to select clean 
data samples of non-interacting CR particles.
In particular, they allow to detect and characterize nuclear interactions occurring at different levels 
in the spectrometers. Two kinds of interaction processes are relevant for the purpose of CR flux measurements: 
(i) nuclear absorption in the top of instrument (TOI) material, and (ii) charge-changing fragmentation processes. 
\begin{figure*}[t]
  \centering
  \includegraphics[width=0.975\textwidth]{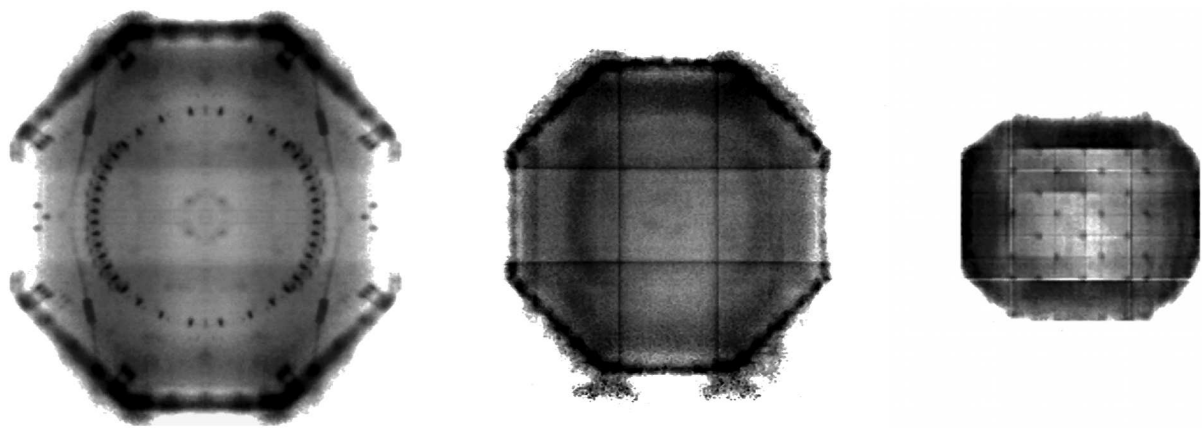}
  \caption{Tomographic reconstructions of the AMS-02 upper material obtained with CR data. The images are generated 
    by measuring the He/p track density ratio at $z=+165$\,cm (left), $z=+132$\,cm (center), and $z=+65.5$\,cm (right).}  
  \label{Fig::ccTACWhite}
 \end{figure*}
Actually both processes may be associated with production of secondary hadrons. 
The case (ii) include those events where secondary particles generated on the TOI are 
triggered and reconstructed, so that they may be confused with genuine CR events.
For both the cases, by the knowledge of the detector material and the relevant interaction cross-sections, 
one can compute the CR survival fractions or theyr probabilities for undergoing charge-changing processes. 
%
It is important to realize that interaction cross-sections are elemental dependent (roughly increasing with 
the mass number), so that elemental ratios such as p/He, B/C, or C/O may need dedicated TOI corrections for interactions.
Further, the CR flux attenuation from different CR species in non-active elements of the TOI material
depends on the effective amount of material along the particles paths, which varies from track to track.
For instance, the flux attenuation of CR protons in the material can be described by ${\rm p}\propto e^{-X/\lambda_{P}}$, 
where $\lambda_{\rm p} \sim$\,85\,g/cm$^{2}$ is the mean attenuation grammage, and it is inversely proportional
to the destruction cross-section $\sigma_{\rm p}$. Similar considerations stand for He and heavier nuclei,
where one has larger interaction probabilities. Roughly, $\lambda_{\rm He}\sim$\,40\,g/cm$^{2}$.
Thus, after traversing an amount $X$ of material, the He/p ratio inside AMS-02 evolves as:
\begin{equation}
{\rm  \left(He/p\right)_{\rm AMS}} \sim {\rm \left(He/p\right)_{\rm CR}} \times e^{-X/X_{0}} \,,
\end{equation}
where (He/H)$_{\rm CR}$ is the impinging CR flux ratio (spatially uniform) and 
$X_{0}\equiv\lambda_{\rm p}\lambda_{\rm He}/\left(\lambda_{\rm p}-\lambda_{\rm He}\right)\sim$\,80\,g/cm$^{2}$. 
Thus, spatial variations on the He/p ratio detected inside AMS-02 can be used to trace 
the presence of TOI material inhomogenities.
Using this idea, we have reconstructed a 3-dimensional map of the AMS-02 material 
with spatial resolution of $\lesssim$\,1\,cm in the $X-Y$ plane and $\sim$\,5\,cm along the $z$\--axis.  
This is shown in Fig.\,\ref{Fig::ccTACWhite}: the images represent a tomographic reconstruction of the AMS-02
material obtained from the He/p track density ratio evaluated at $z=+165$\,cm (above Layer-1), $z=+132$\,cm (inside the TRD), 
and $z=+65.5$\,cm (between the Upper-TOF planes).
To generated these images,  3.7$\times$\,10$^{9}$ $Z=1$ particles and 6.2$\times$\,10$^{8}$ $Z=2$ high-energy nuclei are used 
($R > 2$\,GV), collected during the first year of the mission. 
The darker regions of the image correspond to local He/p ratio deficits which reflect
material inhomogenities of the order of $\Delta X \sim$\,5\,g/cm$^{2}$, 
corresponding to detector elements such as screws, electronics boxes, support structures, and mechanical interfaces. 
\begin{figure}[!t]
  \centering
  \includegraphics[width=0.5\textwidth]{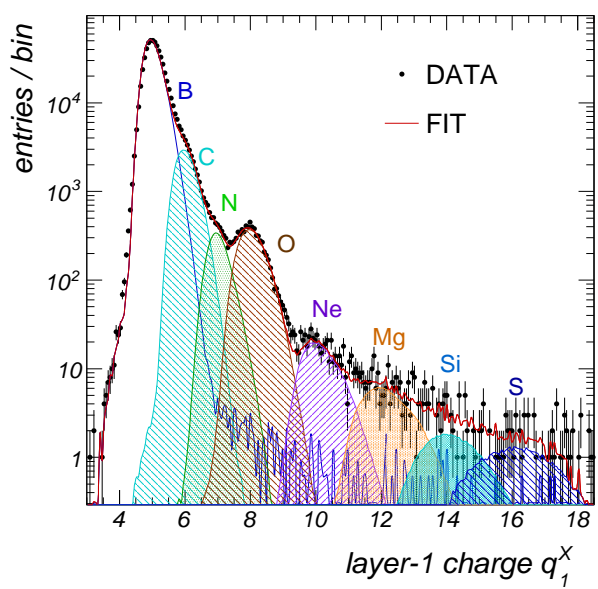}
  \caption{
    Corrected cluster amplitude distribution $q_{1}^{X}$ of the first Tracker layer, in the $X$\--side,
    for CR events identified as Boron with IT+TOF. The figure shows the presence of charge-changing processes 
    occurring in the TRD material, such as C$\rightarrow$B, N$\rightarrow$B, O$\rightarrow$B, etc.
  }
  \label{Fig::ccFragmentationToBoron}
\end{figure}

The second case (ii) appears when incoming $Z$\--charged CR nuclei physically turn into lighter 
fragments ($Z^{\prime}<Z$), after interacting in the TOI material of the spectrometer.
In this case the charge recorded using the IT+TOF system may need to be corrected by the 
knowledge of the probability $P(Z^{\prime}|Z)$.
Since interactions occur mostly in the TRD, the first Tracker layer L1, as well as the TRD itself,
represents a powerful tool for tagging these kind of events. 
Figure\,\ref{Fig::ccFragmentationToBoron} shows the single-layer charge distribution 
observed in L1, after selection $Z^{\prime}=5$ using IT and TOF. 
The colored curves are the single-layer charge responses $q^{X}_{1}$ for known elements.
The data (black markers) are described by a combination of these templates using
Boron ($\sim$\,92\%), Carbon ($\sim$\,5\%), Nitrogen ($\sim$\,0.5\% of N), Oxygen ($\sim$\,1\%) and
heavier elements Ne-Mg-Si-S ($\lesssim$\,1\%). 
These numbers represent the probability $P(Z|Z^{\prime})$ computed with flight data.
In this figure, no quality cuts are applied to the TRD signal.
Clearly, simple cuts on the L1 charge allows to highly-efficiently suppress these events.
Using further criteria on TRD, TOF and IT signals, one may completely reject charge-changing processes below Layer-1 from the event selection.
For instance, the interaction vertices and the charged secondary fragments of the hadronic cascade
may be detected and reconstructed using topological criteria in TRD and Tracker. 
In a similar way, one can also study fragmentation processes occurring in the TOF planes or in lower levels of the detector.
The use of the RICH (see  Fig.\,\ref{Fig::ccAMSDetector}e) may be complementary to the $Z$\--identification procedure, 
given its large dynamical response, especially for high-$Z$ charged nuclei \cite{Bib::ICRC-RICH}.
In Fig.\,\ref{Fig::ccAMSDetector} we show that also the ECAL has charge detection capabilities \cite{Bib::ICRC-ECAL}. 
The ECAL charge estimator $Q$ represents the mean energy deposition per layer. It is computed in the upper 
track segments in the ECAL before the development of hadronic CR showers.

\section{Conclusions} 

We have reviewed the basic principles of charge measurement with the AMS-02 experiment. 
In CR spectra and composition studies, the charge identification is a crucial aspect of the data analysis. 
The procedure relies on the charge response of the IT detectors and the TOF system,
which provide excellent charge detection capabilities for all CR elements. 
Information from other sub-detectors allow to improve the charge separation 
and to detect fragmentation processes occurring in the upper layers of the spectrometer.
All plots shown in this paper are made using flight data. In fact, the large redundancy of 
$Z$-measurements throughout AMS-02 allows to characterize well the charge response of the 
several detector units using the data.


\vspace*{0.5cm}
\footnotesize{{\bf Acknowledgment: }{This work is supported by the \textsf{ENIGMASS} project.}}


\end{document}